\documentclass[aps,prc,floatfix,amsmath]{revtex4-1}
\usepackage{dcolumn}
\usepackage{amssymb}
\usepackage{mathrsfs}
\usepackage{supertabular}
\usepackage{color,graphicx}
\usepackage[bookmarksopen]{hyperref}
\def  \p    {\pi}

\def  \f    {\frac}

\def  \th   {\theta}

\def  \veps {\varepsilon}

\def  \bef  {\begin{figure}}
\def  \eef  {\end{figure}}
\def  \be   {\begin{equation}}
\def  \ee   {\end{equation}}
\def  \ba   {\begin{array}}
\def  \ea   {\end{array}}
\def  \bea  {\begin{eqnarray}}
\def  \eea  {\end{eqnarray}}
\def  \beq  {\begin{eqnarray}}
\def  \eeq  {\end{eqnarray}}
\def  \nn   {\nonumber}
\def  \bd   {\begin{displaymath}}
\def  \ed   {\end{displaymath}}
\def  \bse  {\begin{subequations}}
\def  \ese  {\end{subequations}}
\def  \bwt  {\begin{widetext}}
\def  \ewt  {\end{widetext}}

\def  \ba   {{\bf{a_1}}}

\begin{document}
\title{Non-Fermi liquid correction to the neutrino mean free path and emissivity in neutron star beyond the leading order}
\author{Souvik Priyam Adhya}
\email{souvikpriyam.adhya@saha.ac.in}

\author{Pradip K. Roy}
\email{pradipk.roy@saha.ac.in}

\author{Abhee K. Dutt-Mazumder}
\email{abhee.dm@saha.ac.in}

\affiliation{High Energy Nuclear and Particle Physics Division, Saha Institute of Nuclear Physics,
1/AF Bidhannagar, Kolkata-700 064, INDIA}

\medskip

\begin{abstract}
In this work we have derived the expressions of the mean free path (MFP) and emissivity of the neutrinos by incorporating non-Fermi liquid (NFL) corrections upto next to leading order (NLO). We have shown how such corrections affect the cooling of the neutron star composed of quark matter core. 
\end{abstract}

\maketitle

\section{Introduction}
It has been recently shown that the quantum liquids in the ultra-relativistic regime behaves differently than the normal  Fermi liquid (FL). This is due to the magnetic interaction which becomes important in the relativistic domain and gives rise to the modification of the in-medium dispersion characteristics of the fermions leading to phenomenon unknown in the standard Fermi Liquid theory. The consequence of such modifications have been seen to be very important in determining the thermodynamic properties of ultra-degenerate relativistic matter like specific heat, entropy etc. \cite{holstein73, ipp04, rebhan05, manuel00, sarkar10, sarkar11, tatsumi09}. This in turn, finds application in many astrophysical contexts. For example, inclusion of magnetic interaction modifies the emissivity of neutron stars with quark matter core.  
In this work, we derive the expressions of the neutrino mean free path (MFP) and extend the calculation of emissivity of the neutrinos beyond leading logarithmic approximation. Our results show significant improvement of the previous results \cite{iwamoto82, schafer04, pal11}. One of the qualitative change is the appearance of fractional powers in $(T/ \mu)$ (where T is the temperature and $\mu$ is the chemical potential of the degenerate quark matter) in the expressions of MFP and emissivity as one goes beyond leading order corrections. Subsequently we study the cooling behaviour of the neutron star where the quantitative estimations have been made both for the LO and NLO corrections. We have found out that there is a decrease in the MFP due to NLO corrections compared with the LO case. It is seen that over all corrections to the quantities like MFP and emissivity are significant compared to the Fermi liquid results \cite{iwamoto82, pal11, tubb75, lamb76, adhya12}. 
\section{Formalism}
To calculate the MFP of the neutrinos we consider the simplest $\beta$ decay reactions that occur in the core of neutron star composed of quark matter \cite{shapiro_book},
\begin{eqnarray}
\label{dir}
d+\nu_{e}\rightarrow u+e^-\\
u+e^-\rightarrow d+\nu_{e}.
\label{inv}
\end{eqnarray}
The neutrino MFP is related to the total interaction rate due 
to neutrino emission averaged over the initial quark spins and summed over 
the final state phase space and spins. For the absorption process and it's
inverse, MFP is given by\cite{iwamoto82},
\begin{eqnarray}
\label{mfp01}
\frac{1}{l_{mean}^{abs}(E_{\nu},T)}&=&\frac{g'}{2E_{\nu}}\int\frac{d^3p_d}{(2\p)^3}
\frac{1}{2E_d}\int\frac{d^3p_u}{(2\p)^3}
\frac{1}{2E_u}\int\frac{d^3p_e}{
(2\p)^3 }
\frac{1}{2E_e}
(2\pi)^4\delta^4(P_d +P_{\nu}-P_u 
-P_e)|M|^2 \times\nn\\
&&\{n(p_d)[1-n(p_u)][1-n(p_e)]
+n(p_u)n(p_e)[1-n(p_d)]\},
\end{eqnarray}
where, $g'$ is the total spin and color degeneracies.
In the above expression the squared invariant amplitude
is evaluated as
\cite{iwamoto82} 
$|M|^2=64G^2\cos^2\th_c(P_d\cdot P_\nu)(P_u\cdot P_e)$.
The total emissivity of the non-degenerate neutrinos is obtained by multiplying
the neutrino energy with the inverse of the MFP with appropriate factors and
integrated over the neutrino momentum. The relation is
obtained as\cite{iwamoto82},
\begin{equation}
 \veps=\int \frac{d^{3}p_{\nu}}{(2\pi)^{3}}E_{\nu}\frac{1}{l(-E_{\nu},T)}
\end{equation}
\begin{figure}[htb]
\resizebox{3.5cm}{2.00cm}{\includegraphics{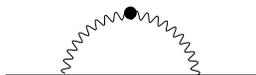}}
\caption{Fermion self-energy with resummed gluon propagator.
\label{fig1}}
\end{figure}
To evaluate the expressions of MFP and emissivity of the neutrinos we will take into account the quark self energy
for the case of degenerate matter.The on-shell self energy of the quarks is severely modified due to interactions within the medium which is manifested in the slope of
dispersion relation for the relativistic degenerate plasma. For quasiparticles
close to the Fermi momentum, the one-loop self energy is dominated by soft gluon
exchanges. For the calculation of  MFP and emissivity, one needs to know the modified
dispersion relation\cite{manuel00,rebhan05},
\begin{equation}
 \omega_{\pm}=\pm(E_{p(\omega_{\pm})} +{\rm
Re}\Sigma_{\pm}(\omega_{\pm},p(\omega_{\pm})))
\end{equation}
where $\omega_{\pm}$ denotes the quasiparticle/antiquasiparticle energy. As we
are
considering only quasiparticles, we will consider only $\omega_+$ and denote it
by
$\omega$. The real part of quark self energy has been determined to be \cite{rebhan05}:
\begin{eqnarray}
&&\rm{Re}\Sigma_{+}(\omega)=-g^2C_Fm\,
 \Big\{{\kappa\over12\pi^2m}\Big[\log\Big({4\sqrt{2}m\over\pi
\kappa}\Big)+1\Big]+{2^{1/3}\sqrt{3}\over45\pi^{7/3}}\left({\kappa\over m}\right)^{5/3}
-20{2^{2/3}\sqrt{3}\over189\pi^{11/3}}\left({\kappa\over m}\right)^{7/3}\nn\\
&&-{6144-256\pi^2+36\pi^4-9\pi^6\over864\pi^6}\Big({\kappa\over m} 
  \Big)^3\Big[\log\left({{0.928}\,m\over \kappa}\right)\Big]
  +\mathcal{O}\Big(\left({\kappa\over m}\right)^{11/3}\Big)\Big\}
\end{eqnarray}
where $\kappa=(\omega-\mu)\sim T$.The NFL effects enter through the modified dispersion relation and will be
required to calculate $dp/d\omega$ needed for the phase space evaluation of the MFP and
emissivity of the neutrinos\cite{pal11,adhya12}. The apperance of fractional power is reminiscent of the NFL characteristic of the self energy \cite{rebhan05}.
\section{Mean free path of neutrinos}
\subsection{MFP of nondegenerate neutrinos}
We now derive MFP for nondegenerate neutrinos 
{\em i.e.} when $\mu_{\nu}\ll T$. Using the free dispersion relation, we obtain the simple FL result \cite{iwamoto82}, 
\bea\label{mfp03}
\frac{1}{l_{mean}^{abs,ND}}\Big|_{FL}=\frac{3C_F\alpha_s}{\pi^4}G_{F}^2\cos^2\th_c
\mu_d\mu_u\mu_e\frac{(E_{\nu}^2+\pi^2 T^2)}{(1+e^{-\beta E_{\nu}})};
\eea
Thus taking into consideration the NFL effects through the phase space modification we obtain at LO \cite{pal11},
\medskip
\bea
\frac{1}{l_{mean}^{abs,ND}}\Big|_{LO}\simeq\frac{C_{F}^{2}\alpha_s}{2\pi^6}G_{F}^2\cos^2\th_c\mu_{e}\frac{(E_{\nu}^2+\pi^2 T^2)}{(1+e^{-\beta E_{\nu}})}
(g\mu)^{2}log\Big(\f{4g\mu}{\pi^{2}T}\Big);
\eea
Extending our calculation beyond the known LO results, we obtain at NLO,
\bea
\frac{1}{l_{mean}^{abs,ND}}\Big|_{NLO}&&\simeq\frac{3C_{F}^{2}\alpha_s}{\pi^4}G_{F}^2\cos^2\th_c\mu^{2}\mu_{e}\frac{(E_{\nu}^2+\pi^2 T^2)}{(1+e^{-\beta E_{\nu}})}\nn\\
&&\times\Big[h_{1}g^{4/3}\Big(\f{T}{\mu}\Big)^{2/3}
+h_{2}g^{2/3}\Big(\f{T}{\mu}\Big)^{4/3}+h_{3}\Big\{1-3log\Big(\f{0.209g\mu}{T}\Big)\Big\}\Big(\f{T}{\mu}\Big)^{2}\Big]
\eea
where the constants are evaluated as,
\bea
h_{1}=0.03;
h_2=-0.149;
h_3=-0.073.\nn
\eea
Similarly, for the scattering of nondegenerate neutrinos in quark matter with appropriate phase space corrections we obtain,
\bea\label{mfp_scnd}
\frac{1}{l_{mean}^{scatt,ND}}\Big|_{FL}&=&\frac{C_{V_{i}}^2 +
C_{A_i}^2}{5\pi}n_{q_i}G_{F}^{2}\f{E_{\nu}^{3}}{\mu};
\eea
\bea
\frac{1}{l_{mean}^{scatt,ND}}\Big|_{LO}\simeq\frac{C_{V_{i}}^2 +
C_{A_i}^2}{30\pi^{3}}n_{q_i}G_{F}^{2}C_{F}\f{E_{\nu}^{3}}{\mu}g^{2}log\Big(\f{4g\mu}{\pi^{2}T}\Big);
\eea
\bea
&&\frac{1}{l_{mean}^{scatt,ND}}\Big|_{NLO}\simeq(C_{V_{i}}^2 + C_{A_i}^2)n_{q_i}G_{F}^{2}C_{F}\Big[l_{1}\f{T^{2/3}g^{4/3}}{\mu^{5/3}}\nn\\
&&+l_{2}\f{T^{4/3}g^{2/3}}{\mu^{7/3}}+l_{3}\Big\{1-3log\Big(\f{0.209g\mu}{T}\Big)\Big\}\Big(\f{T^2}{\mu^3}\Big)\Big],
\eea
where the constants are,
\bea
l_{1}=0.002;
l_{2}=-0.009;
l_{3}=-0.005.\nn
\eea
Thus, the total MFP for non-degenerate neutrinos is obtained by summing up the contributions from the absorption and scattering parts to get the expression of the MFP of the non-degenerate neutrinos up to the NLO terms.
\subsection{MFP of degenerate neutrinos}
This is the case where the neutrino chemical potential ($\mu_\nu$) is considered to be much larger than the temperature, where the neutrinos become degenerate. So, in this case, both the Eq.(\ref{dir}) and reverse Eq.(\ref{inv}) will occur. Using the $\beta$ equilibrium condition and assuming quarks and electrons to be massless, we obtain the following results for the absorption procees,
\bea
\label{mfp_cond1}
\frac{1}{l_{mean}^{abs,D}}\Big|_{FL}=\frac{4}{\pi^{3}}G_{F}^{2}\cos^{2}\th_{c}
\frac{\mu^{2}
\mu_{e}^{3}}{\mu_{\nu}^{2}}\times\Big[1+\frac{1}{2}\Big(\frac{\mu_{e}}{\mu}
\Big)
+\frac{1}{10}\Big(\frac{\mu_{e}}{\mu}\Big)^{2}\Big]
\times[(E_{\nu}-\mu_{\nu})^{2}+\pi^{2} T^{2}];
\eea 
\bea
&&\frac{1}{l_{mean}^{abs,D}}\Big|_{LO}\simeq\frac{2}{3\pi^{5}}G_{F}^{2}C_{F}\cos^{2}\th_{c}\frac{\mu_{e}^{3}}{\mu_{\nu}^{2}}\Big[1+\frac{1}{2}\Big(\frac{\mu_{e}}{\mu}
\Big)
+\frac{1}{10}\Big(\frac{\mu_{e}}{\mu}\Big)^{2}\Big]\nn\\
&&\times[(E_{\nu}-\mu_{\nu})^{2}+\pi^{2} T^{2}](g\mu)^{2}log\Big(\frac{4g\mu}{\pi^{2}T}\Big).
\eea
The NLO result is evaluated as,
\bea
&&\frac{1}{l_{mean}^{abs,D}}\Big|_{NLO}\simeq\frac{8}{\pi^{3}}G_{F}^{2}C_{F}\cos^{2}\th_{c}\frac{\mu_{e}^{3}}{\mu_{\nu}^{2}}\Big[1+\frac{1}{2}\Big(\frac{\mu_{e}}{\mu}
\Big)
+\frac{1}{10}\Big(\frac{\mu_{e}}{\mu}\Big)^{2}\Big]\nn\\
&&\times[(E_{\nu}-\mu_{\nu})^{2}+\pi^{2} T^{2}]\Big[r_{1}T^{2/3}(g\mu)^{4/3}+r_{2}T^{4/3}(g\mu)^{2/3}+r_{3}\Big\{1-3 log\Big(\frac{0.209g\mu}{T}\Big)\Big\}T^{2}\Big]
\eea
Similarly, following the procedure described in \cite{iwamoto82} we obtain,
\bea
\frac{1}{l_{mean}^{scatt,D}}\Big|_{FL}=\f{3}{4\pi}n_{q_i}G_{F}^{2}
[(E_{\nu}-\mu_ { \nu } )^ { 2 } +\pi^{2} T^{2}]\Lambda(x_i);
\eea
\bea
\frac{1}{l_{mean}^{scatt,D}}\Big|_{LO}&\simeq&\f{1}{8\pi^{3}}n_{q_i}C_{F}G_{F}^{2}[(E_{\nu}
-\mu_ { \nu } )^ { 2 } +\pi^{2} T^{2}]\Lambda(x_i)g^{2}log\Big(\f{4g\mu}{\pi^{2}T}\Big);
\eea
\bea
\frac{1}{l_{mean}^{scatt,D}}\Big|_{NLO}&\simeq&\f{3}{2\pi}n_{q_i}C_{F}G_{F}^{2}[(E_{\nu}
-\mu_ { \nu } )^ { 2 } +\pi^{2} T^{2}]\Lambda(x_i)\nn\\
&&\Big[r_{1}g^{4/3}\Big(\f{T}{\mu}\Big)^{2/3}
+r_{2}g^{2/3}\Big(\f{T}{\mu}\Big)^{4/3}+r_{3}\Big\{1-3log\Big(\frac{0.209g\mu}{T}\Big)\Big\}\Big(\f{T}{\mu}\Big)^{2}\Big]
\eea
where the constants are,
\bea
r_1=0.015;
r_2=-0.075;
r_3=-0.036.\nn
\eea
\section{Emissivity of nondegenerate neutrinos}
To calculate the emissivity of the neutrinos we use the expression of the MFP of the nondegenerate neutrinos and obtain,
\bea
\varepsilon  =  \varepsilon_{0}+\varepsilon_{LO} + \varepsilon_{NLO}
\eea
where,
\bea
\varepsilon_{0} \simeq \frac{457}{630}G_{F}^{2}cos^{2}\theta_{c}\alpha_{s}\mu_{e}T^{6}\mu^2
\eea
is the FL result as presented in ref.\cite{iwamoto82}.
At the LO we have obtained,
\bea
\varepsilon_{LO} \simeq \frac{457}{3780}G_{F}^{2}cos^{2}\theta_{c}C_{F}\alpha_{s}\mu_{e}T^{6}\frac{(g\mu)^2}{\pi^2}ln\Big(\frac{4g\mu}{\pi^{2}T}\Big)
\eea
which is in agreement with the result quoted in ref.\cite{schafer04}.
Now, we have obtained the NLO contribution to the neutrino emissivity as,
\bea
\varepsilon_{NLO} \simeq \frac{457}{315}G_{F}^{2}cos^{2}\theta_{c}C_{F}\alpha_{s}\mu_{e}T^{6}\Big[n_{1}T^{2}+n_{2}T^{2/3}(g\mu)^{4/3}-n_{3}T^{4/3}(g\mu)^{2/3}-n_{4}T^{2}ln\Big(\frac{0.656g\mu}{\pi T}\Big)\Big]
\eea
where the constants are evaluated as,
\bea
n_1 = -0.035;
n_2 = 0.015;
n_3 = 0.075;
n_4 = -0.109.\nn
\eea
\section{Results}
\begin{figure}[htb]
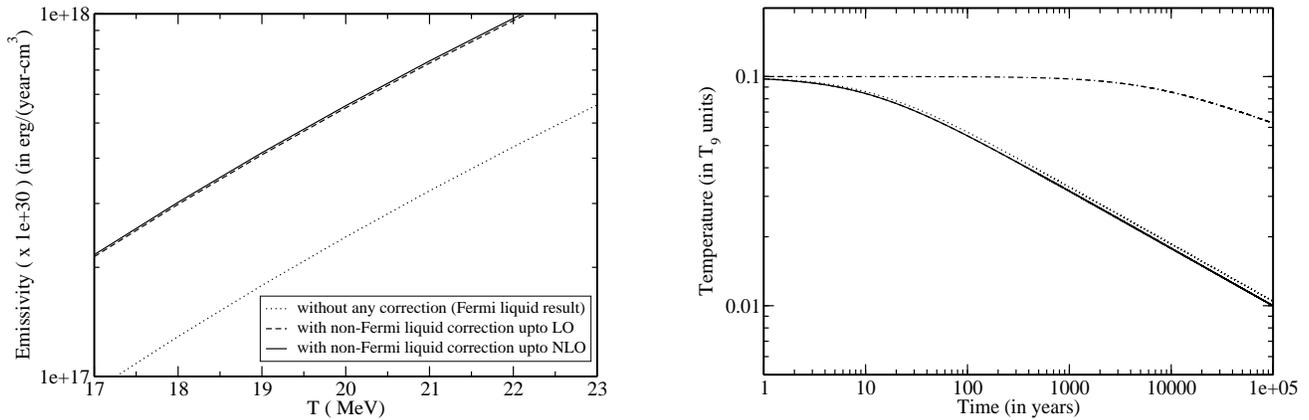

\resizebox{8.0cm}{5.5cm}{\includegraphics{emissivityv1.eps}}
~~~~~~~~~~\resizebox{8.0cm}{5.5cm}{\includegraphics{cooling.eps}}
\caption{The left panel shows the emissivity of the neutrinos with temperature in degenerate quark matter. The right panel shows the cooling behavior of neutron star with core as neutron matter and degenerate quark matter with $T_9$ in units of $10^9$ K. The dotted line represents the FL result, 
the solid line represents the NFL NLO correction. 
The dash-dotted line gives the cooling behavior of  
the neutron star core made up of purely neutron matter.
\label{fig3}}
\end{figure}

\begin{figure}[]
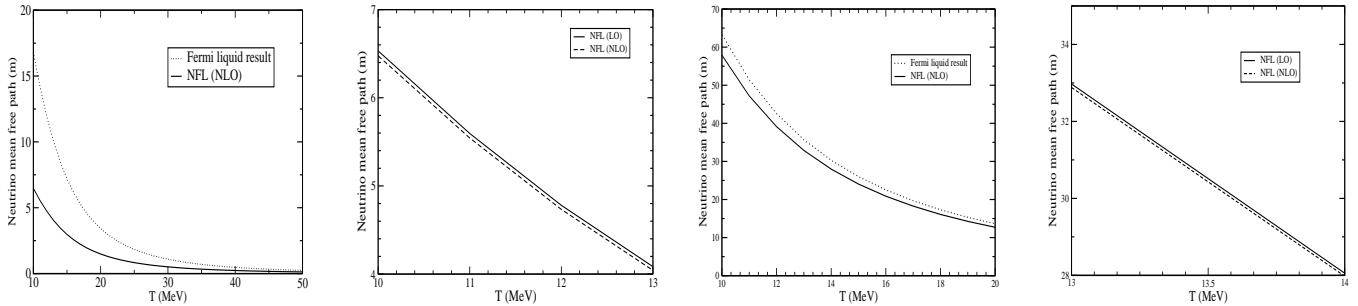

\resizebox{4.0cm}{4.0cm}{\includegraphics{mfp_d.eps}}
~~~~~\resizebox{4.0cm}{4.0cm}{\includegraphics{mfp_d1.eps}}~~~~~\resizebox{4.0cm}{4.0cm}{\includegraphics{mfp_nd.eps}}
~~~~~\resizebox{4.0cm}{4.0cm}{\includegraphics{mfp_nd1.eps}}
\caption{The first figure shows a comparison between the Fermi liquid result and NLO corrections for the NFL effects for degenerate neutrinos. The second figure shows the reduction of the MFP due to NLO corrections over LO results for the degenerate neutrinos. The third and fourth figure shows similar comparisons for the non-degenerate neutrinos.
\label{fig2}}
\end{figure}

In the region of our interest, we assume quark chemical potential of $500$ MeV, electron chemical potential of $15$ MeV and $\alpha_s=0.1$. In fig.(\ref{fig2}) we show that the MFP is decreased due to NFL NLO correction over the LO and Fl case. We find that there is a increase in emissivity of neutrinos due to NLO correction over the LO and FL cases in fig.(\ref{fig3}). The cooling of the neutron star has been numerically studied using the expression of the emissivity of the neutrinos and specific heat of degenerate quark matter upto NLO \cite{rebhan05}. The cooling graph shows faster cooling rate of the neutron star core made up of degenerate quark matter over purely neutron matter. In addition, a comparison has been presented for the NFL NLO and the simple FL case for the case of degenerate matter.
\bigskip
\section{Discussions and Conclusions}
In the present work we find the MFP of both degenerate and nondegenerate neutrinos containing terms which involve fractional powers in $(T/\mu)$ at higher orders. In addition, we have calculated the emissivity of neutrinos and examined NLO corrections over NFL LO and the simple FL case. Finally, we have examined the cooling behaviour of the neutron star involving NLO correction to the emissivity of neutrinos and specific heat of degenerate quark matter. We have found that although there is a modest correction to the quantities like MFP and emissivity of neutrinos over LO and FL case but there is a marginal alteration in the cooling behavior due to such NFL corrections.
\section{Acknowledgments}
One of the authors [SPA] would like to thank UGC, India (Serial No. 2120951147) for providing the fellowship during the tenure of this work.


\begin{thebibliography}{50}
\bibitem{holstein73}
 T. Holstein, R.E. Norton and P. Pincus, Phys. Rev. B {\bf 8}, 2649 (1973).
\bibitem{ipp04} A.Gerhold, A.Ipp and A.Rebhan, Phys.Rev.D {\bf 70}, 105015
(2004); {\bf 69}, R011901(2004).
\bibitem{rebhan05} A.Gerhold and A.Rebhan, Phys.Rev.D {\bf 71}, 085010 (2005).
\bibitem{manuel00} C.Manuel, Phys.Rev.D {\bf 62}, 076009 (2000).
\bibitem{sarkar10} S.Sarkar and A.K.Dutt-Mazumder, Phys.Rev.D {\bf 82}, 
056003 (2010).
\bibitem{sarkar11} S.Sarkar and A.K.Dutt-Mazumder, Phys.Rev.D {\bf 84}, 
096009 (2011).
\bibitem{tatsumi09} K.Sato and T.Tatsumi, Nucl.Phys.A {\bf 826}, 74 (2009).
\bibitem{iwamoto82} N.Iwamoto, Ann.Phys.(N.Y.){\bf 141}, 1 (1982).
\bibitem{schafer04} T.Sch\"{a}fer and K.Schwenzer, Phys.Rev.D {\bf 70},  
114037 (2004).
\bibitem{pal11} K.Pal and A.K.Dutt-Mazumder,Phys.Rev.D {\bf 84}, 034004 (2011).
\bibitem{tubb75} D.L.Tubbs and D.N.Schramm, Astrophys.J.{\bf 201}, 467 (1975).
\bibitem{lamb76} D.Q.Lamb and C.J.Pethick, Astrophys.J.Lett.{\bf 209}, L77 (1976).
\bibitem{adhya12} S.P.Adhya, P.K. Roy and A.K.Dutt-Mazumder, Phys.Rev.D {\bf 86}, 034012
(2012).
\bibitem{shapiro_book} S.L.Shapiro and S.A.Teukolsky, {\it Black Holes, 
White Dwarfs and Neutron Stars.} Wiley-Interscience, New York (1983).
\end{thebibliography}
\end{document}